\documentclass[conference]{IEEEtran}
\IEEEoverridecommandlockouts

\usepackage{cite}
\usepackage{amsmath,amssymb,amsfonts}
\usepackage{graphicx}
\usepackage{textcomp}
\usepackage{xcolor}
\usepackage{booktabs}
\usepackage{microtype}
\usepackage{url}
\usepackage{array}
\usepackage{balance}
\usepackage{tikz}
\usetikzlibrary{shapes.geometric, arrows.meta, positioning, fit, backgrounds}

\begin{document}

\title{Touching and Feeling the Data: A Reusable Software Pipeline for Tactile Statistical Graphs in Accessible Education}

\author{
\IEEEauthorblockN{
Lawrence Obiuwevwi,
Krzysztof J. Rechowicz,
Jessica M. Johnson, 
Erika Frydenlund, \\
Vikas Ashok,
Sachin Shetty,
\& Sampath Jayarathna
}
\IEEEauthorblockA{
\textit{Old Dominion University, Norfolk, VA, USA} \\
lobiu001@odu.edu,
krechowi@odu.edu,
j17johnso@odu.edu,
efrydenl@odu.edu, \\
vganjiqu@cs.odu.edu,
sshetty@odu.edu,
sampath@cs.odu.edu
}
}

\maketitle

\begin{abstract}
Statistical visualization is usually treated as a visual medium, but data can also be touched. Three dimensional printed tactile graphs let blind and low vision students feel distributions, trace trends, and explore relationships through direct haptic interaction. Yet classroom scale use remains limited because producing each graph in CAD software requires specialized skill and hours of manual work. We address this bottleneck as a software problem through a three layer reusable pipeline in about 1500 lines of JavaScript. The first layer derives tactile design parameters automatically from plate dimensions using tactile perception research. The second provides shared chart scaffolding and five modular builders for scatter, bar, histogram, line, and box plots. The optional third layer uses a multi-modal large language model to extract structured chart specifications from uploaded images, with mandatory teacher review before print generation. The pipeline produces print ready binary Standard Tessellation Language files in under 250 milliseconds. We present the design, performance, and limitations.

\end{abstract}


% ============================================================
\section{Introduction}
% ============================================================

For sighted students, a histogram or scatter plot communicates shape, trend, and
outliers almost immediately. For blind or low-vision students \cite{prakash2024understanding,prakash2024towards}, the same chart
often becomes a verbal description delivered by a screen
reader~\cite{prakash2026voxvista,sunkara2025quickque, ferdous2025understanding}, shifting graphical information
into sequential prose and removing the spatial and relational properties that
make charts effective~\cite{Tufte1983,Lundgard2022,Kim2021,Marriott2021}.
Tactile graphics address this gap by turning charts into raised physical forms
that can be explored through touch~\cite{Lederman1987}. FDM-printed charts
encode axes, data features, and Braille at distinct heights, creating a richer
tactile hierarchy than embossed paper~\cite{BANA2010}, and consumer FDM printers
now cost effective abd affordable.

Despite this, we are aware of no systematic classroom deployment of per-lesson
3D-printed statistical graphs. In our engagement with one statistics course
(institution withheld for review), the limiting factor was not printer access
but software: manually modeling each chart in Autodesk Fusion 360 required
about two hours per graph. Existing commercial tools such as TactileView and
ViewPlus IVEO target swell-paper rather than 3D printing and expose no
reusable parameter logic~\cite{TactileView,ViewPlus}. Research systems are
typically one-off artifacts~\cite{Kane2008,Shi2017}.

\begin{figure}[!t]
  \centering
  \includegraphics[width=\linewidth]{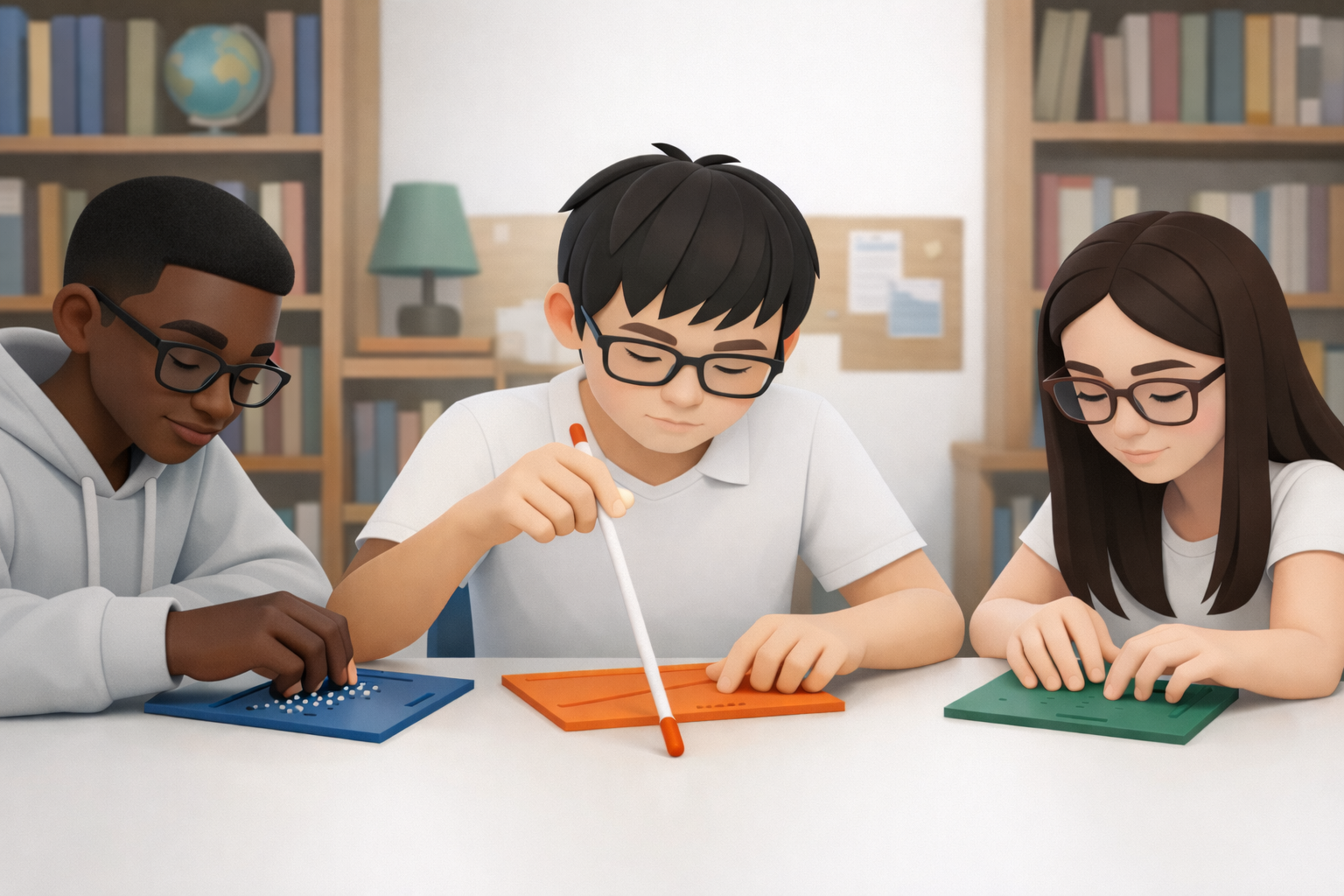}
\caption{Students use reusable 3D-printed tactile plates to explore braille and tactile graphics in an inclusive setting.}
  \label{fig:arch}
\end{figure}

We present a three-layer reusable pipeline that produces a print-ready binary
STL in under one second from plate dimensions, chart type, and numeric data.
Contributions are: \textit{(i)} a parameter derivation layer grounded in
tactile perception research~\cite{Lederman1987,LoC2016}; \textit{(ii)} a
modular geometry layer with five chart builders sharing a common scaffolding;
and \textit{(iii)} an optional vision-assisted extraction layer converting
chart images into editable specifications via a multimodal LLM. To our
knowledge this is the first open-source pipeline that automatically generates
3D-printed tactile statistical graphs from either typed data or chart images,
combining research-grounded parameter derivation with single-pass
Braille-plus-English labeling.

\begin{figure*}[!t]
  \centering
  \includegraphics[
    width=0.95\linewidth,
    height=8.2cm
  ]{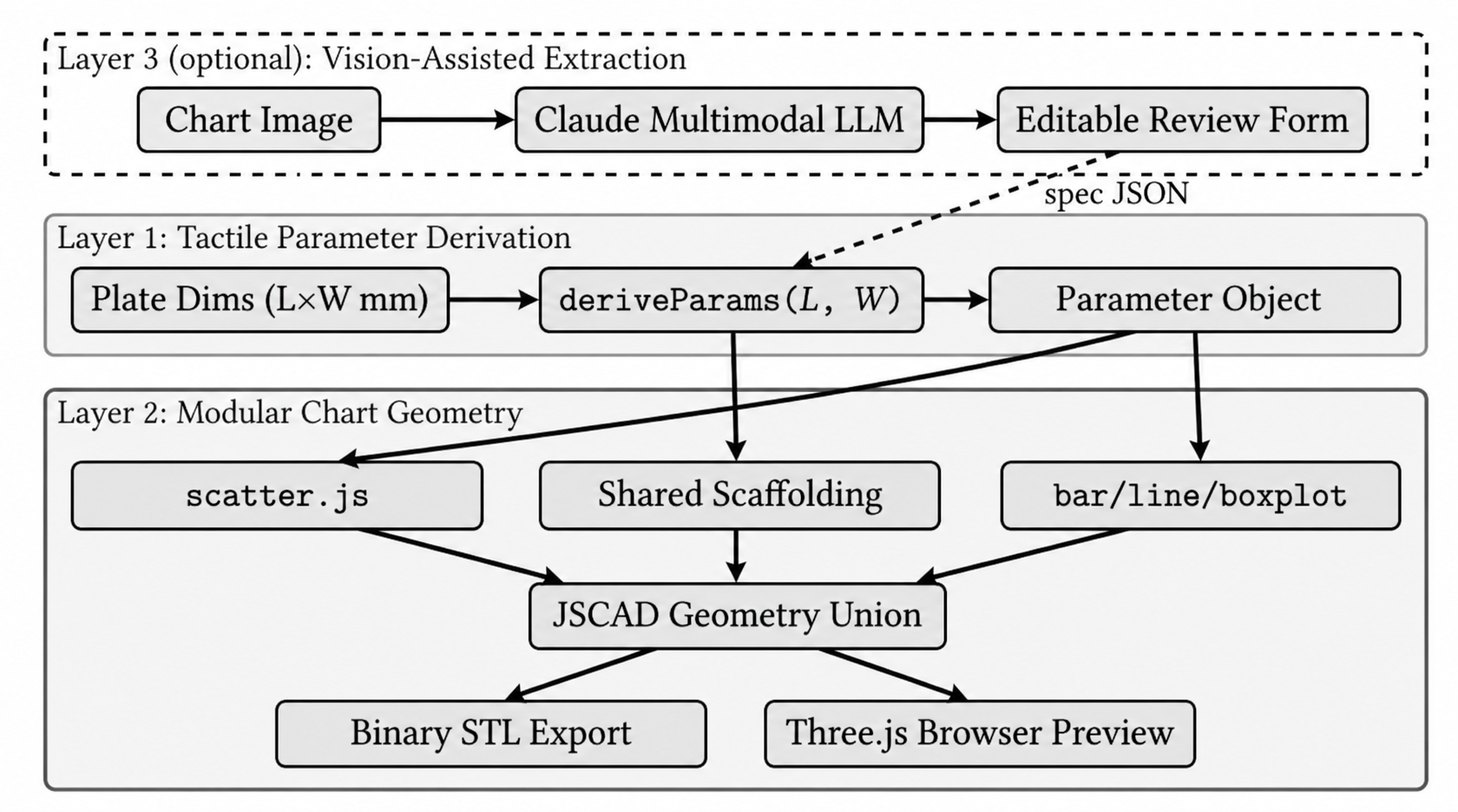}
  \caption{Three-layer reusable pipeline architecture. Layer~3 (dashed, yellow)
  is optional; the core pipeline comprises Layer~1 (tactile parameter derivation,
  blue) and Layer~2 (modular chart geometry, green). Data flows from plate
  dimensions and an optional chart image through to a binary STL file and a
  Three.js browser preview.}
  \label{fig:arch}
\end{figure*}

% ============================================================
\section{Related Work}
\label{sec:related}
% ============================================================

\subsection{Tactile graphics and 3D-printed accessibility}
BANA guidelines~\cite{BANA2010} and Library of Congress
specifications~\cite{LoC2016} define minimum feature heights, line widths, and
Braille dot geometry for fingertip reading; Lederman and Klatzky established
the 0.5\,mm height discriminability floor that anchors our parameter
derivation~\cite{Lederman1987}, and Weinstein's two-point discrimination
findings~\cite{Weinstein1968} motivate the 3.5\,mm scatter-point spacing floor.
Kane et al.\ showed FDM printing can produce usable tactile overlays for
accessibility~\cite{Kane2008}; Shi et al.\ confirmed that properly spaced
raised features significantly outperform verbal descriptions in
comprehension~\cite{Shi2017}; and Hurst and Kane examined barriers to sharing
accessible maker artifacts~\cite{Hurst2013}. Zhao et al.\ highlight the
specific need for automated accessible STEM materials in the
classroom~\cite{Zhao2023}. A common theme is that design logic is embedded in
individual artifacts rather than exposed as reusable software.

\subsection{Chart understanding and accessible visualization}
Academic research on automated tactile chart generation has largely followed the
swell-paper embosser paradigm, focusing on transcription quality rather than
software architecture~\cite{Rowell2003}. On the vision side, Luo et al.'s
ChartOCR~\cite{Luo2021} and Masry et al.'s ChartQA benchmark~\cite{Masry2022}
demonstrated data extraction and question answering over chart images;
Han et al.\ showed instruction-tuned vision models can treat extraction as a
prompt-based task~\cite{Han2023}, and Lee et al.'s VisText advances semantically
rich chart captioning for downstream accessibility~\cite{Lee2023}. Broader
accessibility surveys confirm that multiple modalities are needed to serve
learners with diverse visual needs~\cite{Jung2022}.

\begin{figure*}[t]
  \centering
  \includegraphics[width=\linewidth, height=5.2cm]{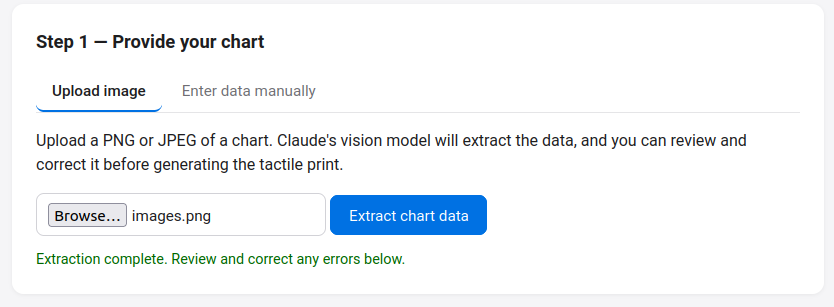}\\[4pt]
  \includegraphics[width=0.48\linewidth, height=6.0cm]{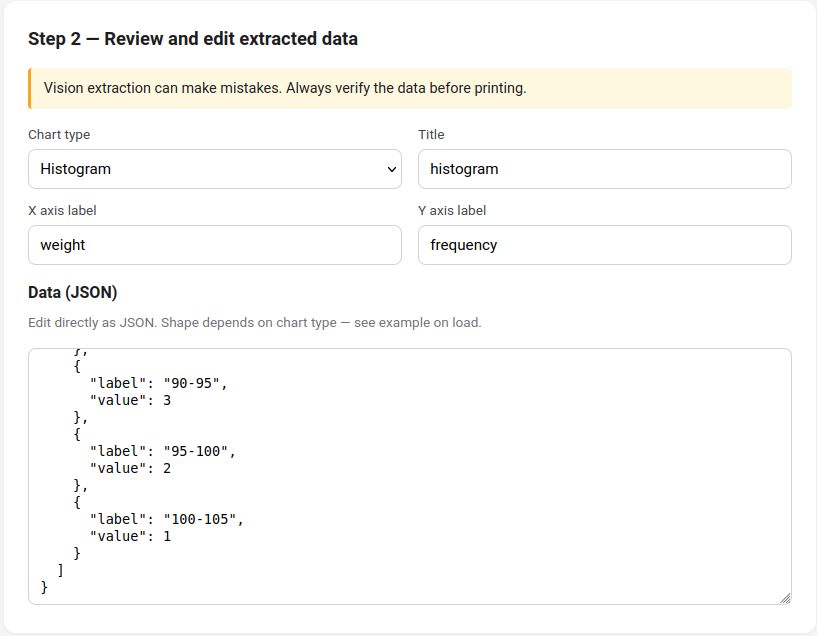}\hfill%
  \includegraphics[width=0.48\linewidth, height=6.0cm]{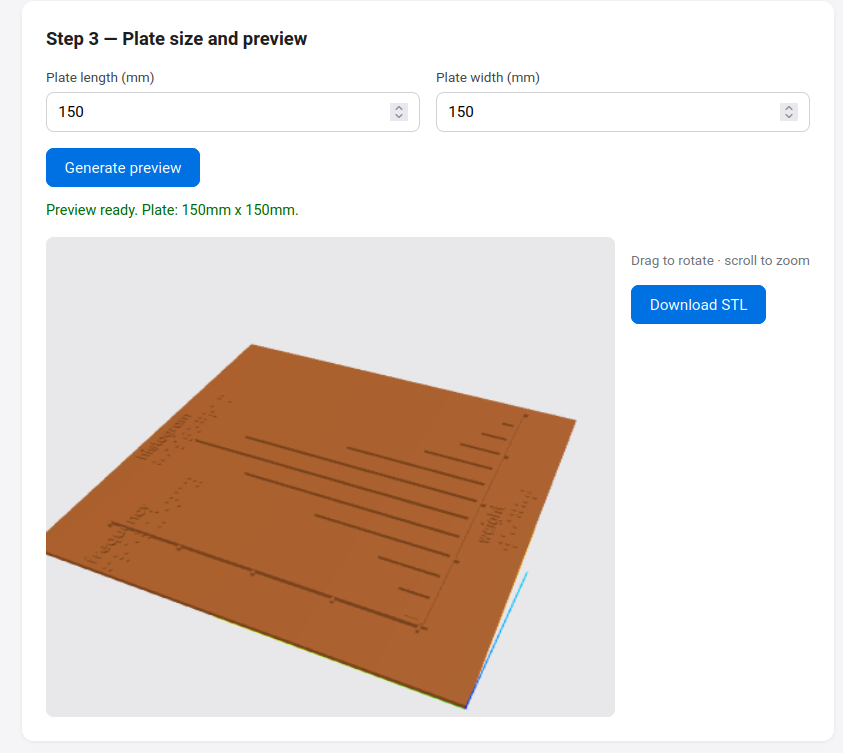}
  \caption{Top: teacher workflow - chart image upload, Claude vision
  extraction, and editable review form. Bottom left: extracted data in the
  review form. Bottom right: Three.js preview of a 150$\times$150\,mm tactile
  histogram plate with raised bars, axis rails, and dual-format labels.}
  \label{fig:ui}
\end{figure*}

% ============================================================
\section{System Design}
\label{sec:design}
% ============================================================

\subsection{Tactile Parameter Derivation}
\label{sec:layer1}

Every chart generation begins with \texttt{deriveParams(L, W)}, which takes
plate length $L$ and width $W$ in millimeters and returns a parameter object
used by all downstream modules. Plate dimensions are clamped to
$[80, 250]$\,mm. Baseplate thickness scales mildly with the smaller dimension:
\begin{equation}
  t_b = \mathrm{clamp}\!\left(\frac{\min(L,W)}{60},\ 2.0,\ 3.5\right)\,\text{mm}
\end{equation}
Plot margins ensure a fixed 18\,mm floor to accommodate dual-format labels
(printed English plus one Braille line), with a plate-proportional
fallback of $0.14\times$ the relevant dimension.

Feature heights are derived from the 0.5\,mm discriminability floor of
Lederman and Klatzky~\cite{Lederman1987}: data features 1.5\,mm
(3$\times$ minimum), axis rails 2.5\,mm ($\approx$1.67$\times$ data features),
scatter/bar/box heights 1.8\,mm, line stroke 1.5\,mm, and Braille dots
0.6\,mm (Library of Congress standard~\cite{LoC2016}: 1.5\,mm diameter,
2.5\,mm intra-cell spacing, 6.0\,mm inter-cell spacing). Axis rails are taller
than data features so a sweeping finger distinguishes structural boundaries
from encoded data. Scatter point minimum separation is 3.5\,mm, exceeding
the 2--3\,mm two-point discrimination threshold~\cite{Weinstein1968}.

\subsection{Modular Chart Geometry}
\label{sec:layer2}

All geometry is generated with JSCAD (\texttt{@jscad/modeling}), running fully
in memory and exporting STL-ready meshes with no graphical environment.

\textbf{Shared scaffolding:} The \texttt{baseGeometry} module creates the
rectangular baseplate, axis rails, and tick marks (1.0\,mm wide, 2.5\,mm
outward, 1.8\,mm tall). Every chart module calls these three builders before
adding chart-specific geometry.

\textbf{Chart modules:} (i)~\textit{Scatter}: point cylinders ($r=1.6$\,mm,
$h=1.8$\,mm); colliding points within 3.5\,mm are merged.
(ii)~\textit{Bar}: width is plot width divided by category count (minimum
4.0\,mm); negative values supported.
(iii)~\textit{Histogram}: bin count set by Sturges'
rule~\cite{Sturges1926}, clamped to $[5,12]$, then rendered through the bar
pipeline.
(iv)~\textit{Line}: rotated cuboid segments (2.0\,mm wide, 1.5\,mm tall) with
vertex cylinders for haptic landmark detection.
(v)~\textit{Box plot}: IQR box from four cuboid walls; median bar
$1.5\times$ IQR wall height; whiskers as 1.2\,mm rail with end caps; outliers
as scatter-point cylinders at 70\% radius.

\textbf{Dual-format labels:} Axis labels and title are printed as JSCAD stroke
English text (6\,mm glyph height, 0.8\,mm raised) above Grade~1 Braille
within the reserved 18\,mm margin. Capital and number indicators are supported;
the vision prompt returns lowercase labels unless capitalization is semantically
necessary, reducing cell count.

\subsection{Vision-Assisted Extraction and Implementation}
\label{sec:layer3}

The optional Layer~3 accepts PNG, JPEG, or WebP chart images and returns a
structured JSON chart specification via a multimodal LLM (Anthropic Claude).
The prompt identifies chart type and outputs typed data: $\{x,y\}$ arrays for
scatter and line charts, label--value pairs for bar and histograms, and
five-number summaries for box plots. Review is mandatory before STL generation.

The server runs on Node.js (v18+) with Express. Geometry is serialized to
binary STL via \texttt{@jscad/stl-serializer}; the frontend provides a Three.js
preview through a native ES module import map with no build step. The codebase
is about 1{,}500 lines across 15 files. The Anthropic API key is required only
for image extraction; the geometry pipeline runs entirely locally.

% ============================================================
\section{Results}
\label{sec:eval}
% ============================================================

\subsection{Pipeline Performance}

Table~\ref{tab:perf} reports STL generation results from \texttt{test-pipeline.js}
on a 150$\times$150\,mm plate. All five chart types complete in under 60\,ms,
well within the 250\,ms threshold perceived as instantaneous~\cite{Miller1968}.
Scatter and box plots have higher primitive counts due to cylinder
tessellation; cuboid-heavy types are faster and smaller.

\begin{table}[h]
  \caption{Pipeline Performance by Chart Type (150$\times$150\,mm plate)}
  \label{tab:perf}
  \centering
  \begin{tabular}{lrrrr}
    \toprule
    Chart type & Primitives & STL (KB) & Triangles & Time (ms)\\
    \midrule
    Scatter   & 214 & 246.6 & 5{,}048 & 51\\
    Bar       & 191 & 170.0 & 3{,}480 & 35\\
    Histogram & 172 & 162.4 & 3{,}324 & 26\\
    Line      & 172 & 181.7 & 3{,}720 & 25\\
    Box plot  & 218 & 199.9 & 4{,}092 & 24\\
    \bottomrule
  \end{tabular}
\end{table}

All five output files satisfy the binary STL formula ($84 + 50T$ bytes exactly),
confirming well-formed output with no truncation. Files open directly in
PrusaSlicer, Bambu Studio, and Cura with no repair required; geometry is
flat-on-bed with no overhangs.

\subsection{Workflow and Extraction Assessment}

The baseline workflow required approximately two hours of Fusion 360 modeling
per chart. The pipeline reduces this to under 60\,ms on-device (excluding
vision API round-trip), a reduction of approximately four orders of magnitude
in software-side labor.

Vision extraction over an informal sample of textbook chart images correctly
identified chart type in all cases and recovered most numeric values within
$\pm 5$--$10\%$ visual estimation error. Two failure modes were observed:
(i)~title-case labels generating unnecessary capital indicators in Grade~1
Braille, causing margin overflow ,  resolved by instructing the model to return
lowercase labels unless semantically necessary; and (ii)~excessive numeric
precision (e.g., $12.3456$ for a visually read value of $12$),  correctable
via the mandatory editable review step before STL generation.

% ============================================================
\section{Discussion and Conclusion}
\label{sec:discussion}
% ============================================================

The layered architecture enables reuse at multiple levels. A researcher can
swap the parameter derivation layer and test its effect across all five chart
types. A developer adding a new chart type need only implement one module;
scaffolding, labels, and STL export are reused automatically. The vision
extraction layer reflects a broader principle: multimodal AI is most useful
here as a \emph{data-entry accelerator}, not as an autonomous producer of
accessible artifacts, making mandatory editable review a safety requirement. %Tactile graphs change the nature of student engagement: rather than receiving verbal descriptions of distributions and relationships, students can trace data directly and build spatial memory of its structure. 
The core finding is that tactile statistical graphics can be produced fast enough and simply enough
for routine classroom use. Future work should conduct formal user studies
measuring student comprehension and teacher task-completion time; add Grade~2
Braille support to reduce label cell counts; integrate with matplotlib and ggplot
figure objects to remove manual data entry; and expand the supported chart set
to violin plots, heat maps, and cumulative distribution functions. %As accessible data representation shifts from accommodation to routine pedagogy, reusable software pipelines of this kind become essential infrastructure.

\section*{Acknowledgement}
This work is supported in part by NSF 245523. Any opinions, findings, and conclusions or recommendations expressed in this material are the author(s) and do not necessarily reflect those of the sponsors.

\bibliographystyle{IEEEtran}
\bibliography{refs}
\end{document}